\documentclass[a4paper,11pt]{article}
\pdfoutput=1 % if your are submitting a pdflatex (i.e. if you have
\usepackage{jheppub} % for details on the use of the package, please
\usepackage{amsmath,amssymb,amsthm,amscd,graphicx}
\usepackage{psfrag}
\usepackage[english]{babel}
\usepackage{float}
\input epsf.sty

\usepackage{color}

\theoremstyle{definition}

\newcommand{\CA}{{\cal A}}

\newcommand{\CF}{{\cal F}}

\newcommand{\CN}{{\cal N}}

\newcommand{\CZ}{{\cal Z}}

\newcommand{\fad}{\operatorname{\Phi}_{\mathsf{b}}}

\newcommand{\im}{{\mathsf{i}}}

\def\IN{{\mathbb N}}
\def\IZ{{\mathbb Z}}
\def\IR{{\mathbb R}}
\def\IC{{\mathbb C}}
\def\IP{{\mathbb P}}

\newcommand{\re}{{\rm e}}
\newcommand{\ri}{{\rm i}}
\newcommand{\rd}{{\rm d}}

\newcommand{\be}{\begin{equation}}
\newcommand{\ee}{\end{equation}}
\newcommand{\ba}{\begin{aligned}}
\newcommand{\ea}{\end{aligned}}
\newcommand{\ben}{\begin{eqnarray}\displaystyle}
\newcommand{\een}{\end{eqnarray}}

\newdimen\tableauside\tableauside=1.0ex
\newdimen\tableaurule\tableaurule=0.4pt
\newdimen\tableaustep
\def\phantomhrule#1{\hbox{\vbox to0pt{\hrule height\tableaurule width#1\vss}}}
\def\phantomvrule#1{\vbox{\hbox to0pt{\vrule width\tableaurule height#1\hss}}}
\def\sqr{\vbox{%
  \phantomhrule\tableaustep
  \hbox{\phantomvrule\tableaustep\kern\tableaustep\phantomvrule\tableaustep}%
  \hbox{\vbox{\phantomhrule\tableauside}\kern-\tableaurule}}}
\def\squares#1{\hbox{\count0=#1\noindent\loop\sqr
  \advance\count0 by-1 \ifnum\count0>0\repeat}}
\def\tableau#1{\vcenter{\offinterlineskip
  \tableaustep=\tableauside\advance\tableaustep by-\tableaurule
  \kern\normallineskip\hbox
    {\kern\normallineskip\vbox
      {\gettableau#1 0 }%
     \kern\normallineskip\kern\tableaurule}%
  \kern\normallineskip\kern\tableaurule}}
\def\gettableau#1{\ifnum#1=0\let\next=\null\else
\squares{#1}\let\next=\gettableau\fi\next}

\newcommand{\figref}[1]{Fig.~\protect\ref{#1}}
\title{\boldmath New results in $\CN =2$ theories from non-perturbative string}

\author{ Giulio Bonelli$^a$, Alba Grassi$^b$ and Alessandro Tanzini$^a$}

\affiliation{
$^a$ International School of Advanced Studies (SISSA), \\
via Bonomea 265, 34136 Trieste, Italy and INFN, Sezione di Trieste\\
\\
$^b$International Center for Theoretical Physics,\\
 ICTP, Strada Costiera 11, Trieste 34151, Italy 
 and INFN, Sezione di Trieste\\}

\emailAdd{bonelli@sissa.it, agrassi@ictp.it, tanzini@sissa.it}
\preprint{
\begin{flushright}
SISSA  14/2017/FISI-MATE \\
\end{flushright}
}

\abstract{We describe the magnetic phase of  $SU(N)$ $\CN=2$ Super Yang-Mills theories in the self-dual $\Omega$ background in terms of a new class of multi-cut matrix models.  These arise from 
a non-perturbative completion of topological strings in the dual four dimensional limit which engineers the gauge theory in the strongly coupled magnetic frame.  
The corresponding spectral determinants provide natural candidates for the $\tau$-functions of isomonodromy problems
for flat spectral connections associated to the Seiberg-Witten geometry.
  }

\begin{document}
\maketitle

\flushbottom

\section{Introduction}

Since the seminal work of Seiberg and Witten (SW) \cite{sw2} a lot of progress has been made to understand $\CN=2$ gauge theories in four dimensions. 
A crucial progress has been obtained by applying equivariant localisation methods to the supersymmetric path integral coupled to the so-called $\Omega$-background \cite{n, Flume:2002az, Bruzzo:2002xf}. This reduces it to a combinatorial expression which has been subsequently linked to quantum integrable systems \cite{ns} and to two-dimensional Conformal Field Theory \cite{agt}.
Localisation methods have been applied so far in the weak coupling limit of the gauge theory rebuilding the low-energy effective field theory of SW in the electric
polarization of the special K\"aehler manifold of Coulomb vacua. More recently \cite{gil1,ilt1,blmst}  it has been realized that the $SU(2)$  Nekrasov and Okounkov (NO) \cite{no2} partition functions in the self--dual  $\Omega$-background are $\tau$-functions of isomonodromy problems related to the corresponding SW geometry, realized as the spectral curve of Hitchin's integrable system.
In this case one can use the well-known relation between isomonodromic deformation problems and Painlev\'e equations to reduce
the evaluation of the NO partition function of gauge theories at strong coupling to the calculation of the relevant $\tau$-function in the long-distance expansion \cite{blmst}. 

On the other hand, four dimensional gauge theories can be engineered by using topological string theory \cite{kkv}. Actually, the latter is richer, in particular at the non-perturbative level there are new effects arising from the embedding of the gauge theory in string theory. Along this line of research a lot of work has been done during the last decade to understand topological string beyond perturbation theory starting with the seminal works \cite{mmopen, Marino:2007te, mmnp}.
 In particular in \cite{ghm},  in the spirit of large $N$ dualities,  a non--perturbative formulation of topological string on toric Calabi-Yau (CY) has been proposed. This formulation has proved to be extremely rich and constructive leading to several new results and applications in various related fields such as integrable systems \cite{hm,Franco:2015rnr,Marino:2016rsq}, supersymmetric gauge theories \cite{bgt, hel} and condensed matter \cite{Hatsuda:2017zwn, Hatsuda:2016mdw}. 
The non-perturbative  proposal of \cite{ghm}  was originally formulated only for CYs whose mirror curve has genus one but it has been  extended to higher genus mirror curves in \cite{cgm2}.

In \cite{bgt} a link between this non-perturbative completion of topological string and isomonodromy problems arising from four dimensional gauge theories has been found in the special case of $SU(2)$ Super Yang-Mills (SYM). The relevant isomonodromy problem in this case is the one associated to Painlev\'e $\rm III_3$ (also called 
$ \rm III^{D_8}$) equation whose $\tau$-function is known since a long time \cite{zamo} to admit a Fredholm determinant description.
Upon a suitable four dimensional scaling limit, the non-perturbative completion of topological string has been shown to be directly related to the Fredholm determinant above. This produces a matrix model presentation of the gauge theory partition function in the strongly coupled magnetic frame  and provides an operator theory interpretation of the self--dual $\Omega$ background  ($\epsilon_1=-\epsilon_2=\epsilon$). 
Moreover, it also provides exact S-duality transformation formula for the Nekrasov partition function with self-dual $\Omega$-background including non-perturbative corrections
in $\epsilon$.

The purpose of this paper is to extend these results to $SU(N)$ gauge theories. More precisely, in section \ref{su2rev} we review the  consequences of the genus one proposal for $SU(2)$ theories  and we provide the exact S-duality transformation for the corresponding gauge theory partition function. In section \ref{nps}, by following the general prescription of \cite{cgm2},  we derive the matrix models computing the topological string partition function on the $Y^{N,0}$ geometries.  The result is given by the $N-1$ cut matrix model shown in \eqref{mmsu35d}. Then, in section \ref{sun4}, we perform  the so-called {\it{dual}} four dimensional limit \cite{bgt} on these models and we make contact with $\CN=2$ $SU(N)$ SYM in the four dimensional self-dual  $\Omega $ background \cite{no2}. More precisely we find that the partition function in the magnetic frame is given by 
\be \label{su3mmintro} {\ba Z_{\rm N}^{\rm 4d}(M_1,\cdots,M_{N-1})=&  {1 \over M_1!  \cdots M_{N-1}!} \int {{\rm d} ^M x\over (2\pi)^M} \prod_{j=1}^{N-1} \prod_{i_j\in I_j}\re^{-  \frac{N \Lambda }{ \pi ^2 \epsilon}{ \sin \left(\frac{\pi  j }{N}\right)} \cosh(x_{i_j})} \\
&   \times {\prod_{ 1\leq i<j \leq M}2\sinh\left({x_i-x_j\over 2}+{1\over 2} (d_i-d_j) \right)2\sinh\left({ x_i-x_j \over 2}+{1\over 2}(f_i-f_j)\right) \over \prod_{i,j=1}^M2\cosh\left( {x_i-x_j \over 2} +{1\over 2} (d_i-f_j) \right)},
\ea }\ee
where $\Lambda$ denotes the instanton counting parameter in gauge theory. The shifts
 $f_i, d_i$  are given in  \eqref{mmsu35d} and they depend on the rank $N$ of the gauge group. We also
 used \be I_j= \left[{\sum_{s=0}^{j-1} M_{s}},\sum_{s=1}^{j} M_{s}  \right] \cap \IN , \quad M_0=1, \quad M=\sum_{i=1}^{N-1}M_i . \ee
As a consequence we have a spectral determinant representation for the four dimensional Nekrasov-Okounkov  partition function associated to these $SU(N)$ theories as shown in section \ref{opno}. We  expect this to be the $\tau$-function of the isomonodromy problem associated to the Hitchin's system describing the relevant
SW curve, see the end of section \ref{opno}. The spectral determinant presentation also allows to compute the exact S-duality transformation of the $SU(N)$ Nekrasov partition function.

\section{Reviewing the $SU(2)$ case} \label{su2rev}

The TS/ST duality \cite{ghm} has led to various exact results in topological string and in spectral theory which  allows us to explore all range of the couplings in both side of the duality. We denote by $g_s$ the coupling constant of string theory and by $\hbar\sim g_s^{-1}$ the Planck constant appearing in the spectral theory side of the correspondence. 

It was pointed out in \cite{bgt} that there are two limits in which the TS/ST duality makes contact with gauge theory.  In one limit,  which we refer to as  the {\it standard} four dimensional limit, one takes $\hbar\to 0$. It was shown in \cite{hm} that in this limit the TS/ST correspondence  reproduces  the well known Nekrasov--Shatashvili  (NS) conjecture in four dimensions \cite{ns}. In the second limit instead,  which we refer to as  the {\it dual} four dimensional limit,  one takes $\hbar\to \infty$ and we obtain some new results in $\CN=2$ gauge theories. 
In this section we review these two limits and their consequences for  $SU(2)$ gauge theories.
 Even tough we mainly focus on the  pure $SU(2)$ theory this procedure can in principle be extended as well to other  geometries which engineer $SU(2)$ gauge theories with $N_f$ matter multiplets.

\subsection{Operator theory and the self-dual $\Omega$ background}
It is  known \cite{adkmv,acdkv, ghm, kama,Laptev:2015loa} that quantization of mirror curves to toric CY leads to well defined operators with a positive and discrete spectrum.
One of the simplest examples is the canonical bundle over $\IP^1\times \IP^1$. The quantization of its mirror curve, namely
\be   \label{mirrorp1}  \xi^{1/2} \left(\re^x +\re^{-x} \right)+\re^p +\re^{-p}=\kappa, \quad x, p \in \IC, \ee
 leads to the following operator which corresponds to the Hamiltonian of $SU(2)$ relativistic Toda lattice 
\be \label{rt} {\rm O}= \xi ^{1/2} \left(\re^{\mathsf x}+\re^{-\mathsf x}\right) +\re^{\mathsf p}+\re^{-\mathsf p}, \qquad [\mathsf x, \mathsf p]=\ri \hbar. \ee
For the propose of this paper we take \be \xi, \hbar \in \IR_+. \ee 
 The spectral properties of the operator ${\rm O}$ are determined by a special combination of the NS and  the unrefined (or GV) limit of topological string \cite{nek5,acdkv,Mironov:2009uv,ghm}. 
 More precisely the perturbative WKB part of the spectrum is closely related to the NS limit  as pointed out in \cite{nek5,Mironov:2009uv,Bonelli:2011na,acdkv}. However there are additional non--perturbative corrections which are encoded in the  GV limit of topological string \cite{km,ghm}  \footnote{In this paper we are  interested in the spectral determinant  formulation of the TS/ST duality in which the GV part plays a crucial role. However if one focus exclusively on the quantization condition, by using blowup equations it is possible to rewrite the GV part of the quantum volume by using the NS free energies  in a way which features S-duality \cite{wzh,hm,Franco:2015rnr,huang1606,Grassi:2016nnt}. Recently  this relation  has lead to an extension of the blowup formalism \cite{Gu:2017ccq}.}.
   In the following we will describe two different limits of the operator  \eqref{rt} dominated respectively by one of the two contributions.

\subsubsection{The standard four dimensional limit}
 
 In the standard four dimensional limit  \cite{nek5,kkv,ikp3,hm}
%In the case of
 one  scales the parameters  of  the operator \eqref{rt} as
 { \be \label{s4} \xi= \beta^{-4}, \quad \hbar=\beta \hbar_{\rm  T}, \quad \mathsf x = \mathsf  {\overline x  }\beta \qquad \beta \to 0.\ee}
This leads to
\be {\rm O} \quad \xrightarrow{\beta \to 0} \quad {\rm H_{\rm 4D}}\ee
where ${\rm H}_{\rm 4D}$ is the Hamiltonian of quantum Toda namely
{ \be {\rm H}_{\rm 4D}=\re ^{\mathsf p}+\re^{- \mathsf p}+\mathsf{ {\overline x  }} ^2 \quad [\mathsf {\overline x}, \mathsf  { p  }]=\ri\hbar_{\rm T}.\ee}
The eigenvalues of this Hamiltonian are  \be E_{\rm 4D}^{n}=\lim_{\beta \to 0}\left(\re^{E_n}-2\beta^{-2}\right),\ee
where $\re^{E_n}$ denotes the energies of \eqref{rt}.
As conjectured in \cite{ns}, the spectrum of  $\rm H_{\rm 4D}$ is exactly determined by the NS limit of $\CN=2$ $SU(2)$  SYM  theory. Indeed in this standard 4d limit  the additional non-perturbative corrections appearing in the relativistic version of the operator, and computed in \cite{ghm}, are not present as pointed out in \cite{hm}. 

At the level of  flat coordinates this limit can be implemented  by taking the so--called geometric engineering limit \cite{kkv}
{ \be \ba  T_b=- \log \left(\beta ^4\right)+\beta L, \quad
T_f=\beta L, \quad
g_{st}=4 \pi^2 \beta^{-1} \hbar_{\rm T}^{-1}, \quad \beta \to 0 \ea\ee}
where $T_b , T_f$ are the Kahler parameters of the canonical bundle over $\IP^1\times \IP^1$ .

 \subsubsection{The dual four dimensional  limit} \label{dus}
As discussed above in the standard four dimensional limit \eqref{s4} we take 
$ \hbar \to 0$  and
we make contact with $\CN=2$ SYM in the four dimensional NS phase of the $\Omega$ background.
  It is therefore natural to ask  whether there is a different limit of the relativistic operator taking
\be \hbar \to \infty  \ee
and still making contact with a different phase of  four dimensional $\CN=2$ theories. In \cite{bgt} it was pointed out that this limit  indeed exists,   we refer to it as {\it dual four dimensional limit}. In the case of local $\IP^1 \times \IP^1$ this consists in taking 
\be \label{dl} \ba &  \hbar= { 1 \over \epsilon \beta}, \quad \log \xi =   {a\over 2 \pi} -\frac{  \log \left(\beta ^4 \Lambda^4 \right)}{ 2 \pi \beta  \epsilon }, \\
& \log \kappa=- {\hbar \over 4 \pi} \log \left(\beta^{4} \Lambda^{4}\right)+\log  \left( 1+\re^{  {a\over 2 \pi} }\right)+ \mathcal{O}(\re^{\beta^{-1}\log \beta^4}), \qquad { \beta \to 0^+} , \ea \ee
 where $\kappa$ is the complex modulus of  local $\IP^1\times \IP^1$ and, after quantization, it is related to the  eigenvalues of the operator \eqref{rt}. From the four dimensional perspective the parameter $a$
is related to the A period of the $SU(2)$ SW curve \footnote{ The parameter $a$ is related to the SW variable $\tt a$ as $a \sim {\tt a / \epsilon}$.},   $\Lambda$ is the instanton counting parameter and  $\epsilon$ is the twisting parameter of the self-dual $\Omega$ background. 
 In general, to implement the dual limit on the operator \eqref{rt} it is more laborious  than the standard four dimensional limit but it can be done  as explained in \cite{bgt}.  In particular one has to use unitary transformations to write the operator in  a suitable form.
In the case  of local $\IP^1 \times \IP^1$ one has to rescale $\mathsf x, \mathsf p$ as
 \be \mathsf x=\mathsf v {1\over 2}+\mathsf u {\hbar \over 2 \pi }+{1\over 4 } \log \xi , \quad  \mathsf p=\mathsf v{1\over 2}-\mathsf u {\hbar \over 2 \pi }+{1\over 4 } \log \xi , \quad  [\mathsf u , \mathsf v ]= 2 \pi \ri .\ee
 By using unitary transformations, one can write \eqref{rt} as \cite{kmz}
\be \label{uvt} \xi ^{1/4} | g\left(b{\mathsf{u}\over 2 \pi }\right) | { 2\cosh(\mathsf{v}/2)} | g\left(b{\mathsf{u}\over 2 \pi }\right)|,\ee
where \be  g (\mathsf{s}): =\re^{-\pi b \mathsf{s}/2}{  \fad(\mathsf{s}+\log \xi /(4 \pi b)- \ri b/4)\over \fad(\mathsf{s}-\log \xi /(4 \pi b)+ \ri b/4) }, \quad  b^2={\hbar \over \pi },
\ee
and  $\fad $ denotes the Faddeev's quantum dilogarithm \cite{faddeev,fk}.
Once the operator  is written in  the form \eqref{uvt} it is easy to implement the dual limit \eqref{dl}. As explained in  \cite{bgt}  one obtains, after a suitable normalisation,  the following operator 
\be \label{rho4d} {\rm O}_{\rm 4D}= \re^{{ \pi^{-2}  \Lambda \epsilon^{-1}  } \cosh(\mathsf u)}{ \left(\re^{\mathsf v/2}+ \re^{-\mathsf v/2} \right)}\re^{{\pi^{-2}  \Lambda \epsilon^{-1} } \cosh(\mathsf u)}, \qquad [\mathsf u, \mathsf v]= 2 \pi \ri  .\ee 
The spectrum of  this operator is exactly computed by the  $\CN=2$ $SU(2)$  SYM  in the self--dual $\Omega$ background.  Therefore the dual limit provides an operator interpretation of the self-dual $\Omega$ background.

At the level of  flat coordinates  this limit consists in taking
\be \ba  T_b =-{1\over 2 \pi \beta \epsilon}\log \left(\beta^4 \Lambda^4\right) \quad
T_f =-{ a\over 2\pi}, \quad
g_{st}=4 \pi^2 \epsilon \beta, \quad \beta \to 0 \ea\ee
where $T_b , T_f$ are the Kahler parameters of the canonical bundle over $\IP^1\times \IP^1$. This can be viewed as a sort of dual (or rescaled) geometric engineering limit where we replaced  \be  T_b,\quad T_f, \quad  g_{st}\quad \to  \quad g_{st}T_b,\quad g_{st}T_f, \quad g_{st}^{-1} .\ee

  \subsection{The TS/ST duality and Painlev\'e equations}
  
  The implementation of the dual limit has also an additional interesting feature: it connects the TS/ST duality to the theory of Painlev\'e equations \cite{bgt}. This goes as follows.

In a series of papers \cite{ilt1,ilt,gil1,ilte, bes,gil,blmst} it was found that the  $\tau$ functions of Painlev\'e equations are computed by four dimensional $\CN=2$, $SU(2)$ gauge theories in the self--dual $\Omega$ background. Different types of Painlev\'e equations correspond to different matter content in the gauge theory. For instance the pure $SU(2)$  theory computes the $\tau$ function of Painlev\'e $\rm III_3$ (also known as $ \rm III^{D_8}$) while the Painlev\'e $\rm V$ is related to  $SU(2)$ gauge theory with $N_f=3$ fundamentals multiplets.

\begin{figure}[h!] \begin{center}
 {\includegraphics[scale=0.35]{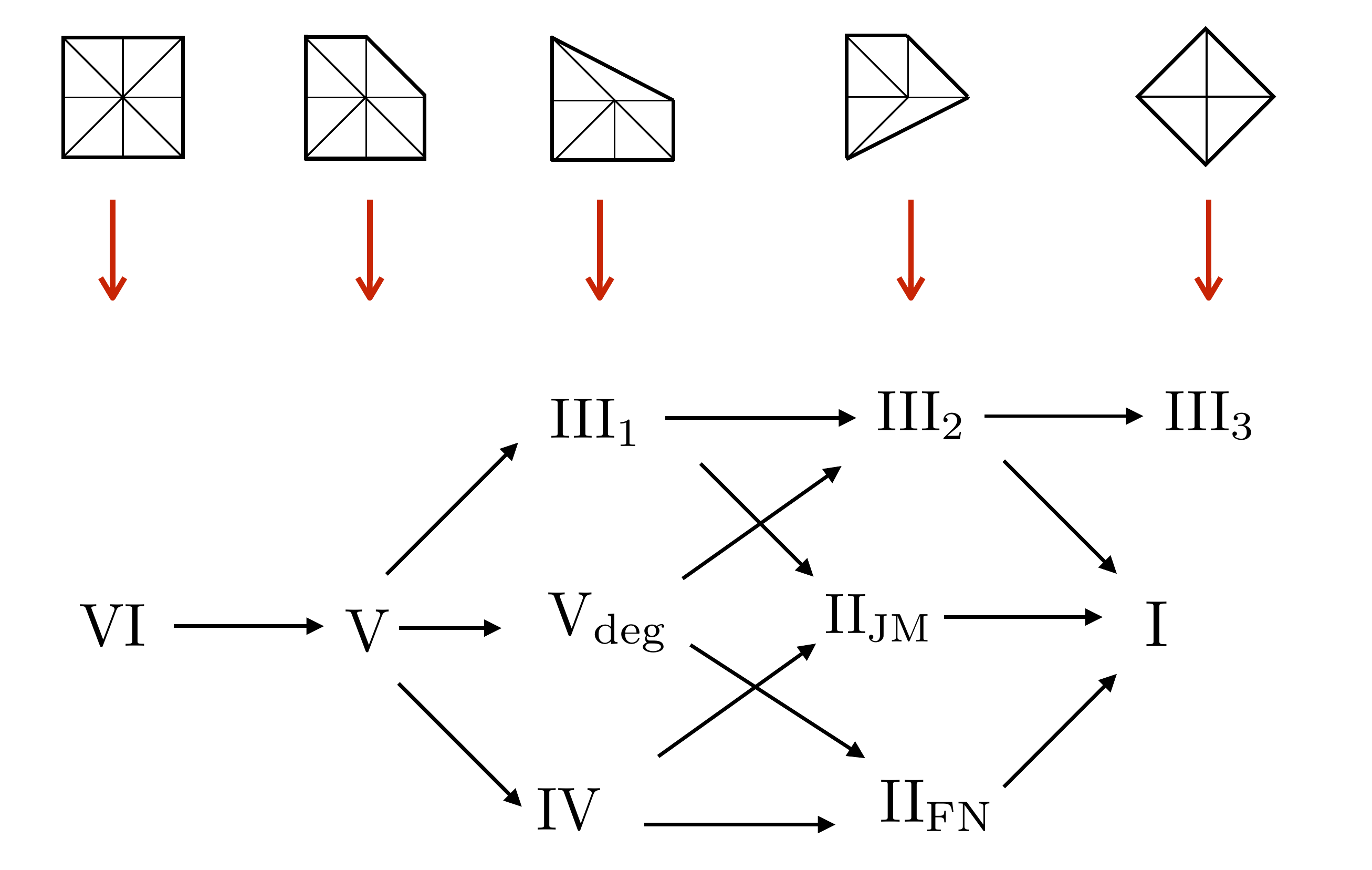}}
\caption{In the upper line we have a list of  polyhedra representing del Pezzo surfaces $S$  \cite{Bat,ckyz} connected to the coalescence diagram of Painlev\'e equations through an arrow. The total space of the canonical bundle  over $S$ is a CY manifold and we can put topological string theory on it. In particular  we can consider the TS/ST duality for these manifolds. Once we implement the dual 4d limit on the determinants appearing in this duality  we recover the tau function of a corresponding Painlev\'e equation.  Also notice that there may be different del Pezzo's which engineers  the same Painlev\'e equation. In the figure we have chosen one of them \cite{Eguchi:2000fv}. } \label{fig1}
\end{center}
\end{figure}

From the string theory viewpoint  we expect that  these results can be obtained by implementing the dual four dimensional limit to the TS/ST duality  on some specific CY  $X$ with genus one mirror curve \cite{bgt}.
 In this duality one associates  a trace class operator $\rho_X$ to any of these geometries and gives an exact expression for the corresponding spectral determinant.  Schematically we have \cite{ghm}
\be \label{det5d}\det (1+\kappa \rho_X)=\sum_{n\in \IZ} \re^{J_X(\mu+2\pi \ri n, \hbar)}, \quad \kappa=\re^{\mu} \ee  
where  $J_X$ is the topological string grand potential studied in \cite{mp,hmo2,cm,hmo3,hmmo,ghm}. If $X$ is the  canonical bundle  over $\IP^1\times \IP^1$ the operator $\rho_{\rm \IP^1\times \IP^1}$ is the inverse of \eqref{rt}.
When the underling geometry $X$ can be used to engineer gauge theories \cite{kkv,ta,ikp3,Eguchi:2000fv} we can implement the dual four dimensional limit described above. After performing this limit  \eqref{det5d} takes the following form
\be \label{det4d} \ba & \det (1+\kappa \rho_{X}^{\rm 4D})=\sum_{n\in \IZ} \CZ_X^{\rm 4D}{(\sigma+ n, t)},\\
& \quad   t\sim \Lambda^4/\epsilon^4, \qquad  \sigma \sim a , \quad \kappa \sim \cos( 2  \pi\sigma)\ea\ee
where $a, \epsilon, \Lambda$ are  the gauge theory parameters introduced in section \ref{dus}.   In the  context of Painlev\'e equations $t$ is the time and the variable $\sigma$ is related to the asymptotic conditions. We  denote  by $\rho_{X}^{\rm 4D}$ the operator obtained after performing the dual limit on $\rho_{X}$. 
Moreover, up to an overall factor, $ \CZ_X^{\rm 4D}$ is the partition function of  a corresponding $\CN=2$ $SU(2)$ SYM theory in the four dimensional self--dual background. The specific matter content of the gauge theory depends on the starting geometry $X$ according to the geometric engineering construction \cite{kkv,ta,ikp3,Eguchi:2000fv}.  
In particular on the r.h.s of \eqref{det4d} we recover, up to an overall factor, the $\tau$ function of a corresponding Painlev\'e  equation as computed in \cite{ilt1,ilt,gil1,gil,blmst} \footnote{\label{fn3}The general form for the $\tau$ function of Painlev\'e equations  depends in general on two variables $\sigma, \eta$ which are related to the monodromy data of the corresponding Fuchsian system. To make contact with TS/ST we set $\eta=0$ \cite{bgt}. As explained in \cite{mmnp,fki3} fixing $\eta$   corresponds to fix the non perturbative ambiguity. Hence the TS/ST duality  corresponds to one particular non-perturbative choice. }.   On the l.h.s.~instead we obtain the spectral determinant representation of the same $\tau $ function.
Hence from this perspective  different Painlev\'e equations correspond  to different background geometry in string theory as illustrated on \figref{fig1}.
Therefore  this construction gives a geometrical meaning to the operators whose spectral determinant compute the $\tau$ function of Painlev\'e equations: these arise by quantizing mirror curves to  CY geometries and performing the dual limit.
 The details of this limit have been worked out for pure $SU(2)$ in \cite{bgt} which makes contact with the local $\IP^1\times  \IP^1$ geometry and the Painlev\'e $\rm III_3$ equation.   In this case the equation \eqref{det4d} reads
 \be  \label{4dstat}\det (1+\kappa \rho_{\IP^1 \times \IP^1}^{\rm 4D})=\re^{\frac{\log (2)}{12} +3\zeta'(-1)}t^{-1/16}\re^{4 \sqrt{t}} \sum_{n \in \IZ} Z^{\rm Nek}(\sigma+n, t),   \ee
 where we used \be \label{dic4}  \kappa={\cos \left(2 \pi \sigma\right)\over 2 \pi} , \quad  t= \left(\frac{\Lambda }{4 \pi ^2 \epsilon }\right)^4\ee
  and $\rho_{\IP^1 \times \IP^1}^{\rm 4D}$ is the inverse of \eqref{rho4d}. More precisely
 \be  \rho_{\IP^1 \times \IP^1}^{\rm 4D}= \re^{-{4 t^{1/4} } \cosh(\mathsf u )}{4 \pi  \over \left(\re^{\mathsf v/2}+ \re^{-\mathsf v /2} \right)}\re^{-{4 t^{1/4} } \cosh( \mathsf u)}, \qquad [\mathsf u,  \mathsf v]= 2 \pi \ri .\ee
Moreover we denoted by $Z^{\rm Nek}(\sigma, t)$ the Nekrasov partition function of pure $\CN=2$, $SU(2)$ theory, namely  
  \be \label{nek4d}\ba Z^{\rm Nek}(\sigma, t)={t^{\sigma^2}  \over  G(1-2\sigma)G(1+2\sigma)} \left( 1+\frac{t}{2 \sigma ^2}+ \frac{\left(8 \sigma ^2+1\right) t^2}{4 \sigma ^2 \left(4 \sigma ^2-1\right)^2}+\mathcal{O}(t^3)\right). \ea\ee
  Sometimes we refer to
  \be Z^{\rm NO}(\sigma, \Lambda, \epsilon)= \sum_{n \in \IZ} Z^{\rm Nek}(\sigma+n, t)\ee
  as the Nekrasov-Okounkov  partition function.  
 The statement \eqref{4dstat} has been proved  for   $\sigma \neq \IZ/2$  in \cite{bgt} by using the relation with the Painlev\'e $\rm III_3$ equation.
 
 Even tough the details have been worked out only for the pure $SU(2)$ gauge theory we expect that a straightforward generalisation should be possible for the others geometries illustrated on \figref{fig1} as well. 

We would like to observe that this approach only holds when some particular initial conditions are imposed on the solutions to Painlev\'e equations (see footnote \ref{fn3}). Recently a more general formalism to compute generic  Fredholm determinant representations of isomonodromic tau functions has been developed in  \cite{lnew}.  It would be interesting to derive the kernels $\rho_{X}^{\rm 4D}$ as limiting cases of this general formulation.

\subsection{Dual matrix models in four dimensions }\label{newsection}
In addition the TS/ST duality can be used to compute the exact expression for the partition function of topological string theory on $X$. Let us review how this goes. By using standard results in Fredholm theory we have
\be \det (1+\kappa \rho_{X})=\sum_{N\geq 0} \kappa^N Z_X(N), \ee
where $Z_X(N)$ are the fermionic spectral traces of $\rho_{X}$, namey
\be \label{mm} Z_X(N)={1\over N!}\sum_{\sigma \in S_N}(-1)^{\sigma} \int_{\IR} \rd ^N x  \rho_{X}(x_i, x_{\sigma(i)}).\ee
We denoted by $S_N$  the permutation group of $N$ elements.  The TS/ST duality states that $Z_X(N)$ is the non-perturbative partition function of topological string on $X$. More precisely, at the level of the fermionic spectral traces the equality \eqref{det5d} reads \cite{ghm}
\be \label{zj}Z_X(N)={1\over 2 \pi \ri} \int_{\mathcal{C}}  \re^{J_X(\mu, \hbar)- \mu N}\rd \mu ~ ,\ee
where $\mathcal{C}$ is the Airy contour.  Furthermore by using the Cauchy identity it is possible to write \eqref{mm} as a matrix model which computes the partition function of  topological string in the conifold frame \cite{mz}.    Even tough $Z_X(N)$ is originally defined in \eqref{mm} for integers $N$, it was pointed out in \cite{cgm}  that  \eqref{zj} provides  an analytic continuation of the fermionic spectral traces to an entire function in the full complex plane. 

When we implement the dual limit at the level of the fermionic spectral traces  \eqref{mm},  we obtain a matrix model expression for the partition function of  $d=4$, $\CN=2$, $SU(2)$ gauge theories  in the self-dual $\Omega$-background in the magnetic frame.  This was done in details for the pure $SU(2)$ theory in \cite{bgt} where it was found that the matrix model computing its partition function is a well known $O(2)$  model
\be Z_2^{\rm 4d}(M)= {1\over M!} \int \prod_{i=1}^M  \frac{\rd x_i}{4 \pi } \re^{-{1\over g_s }\cosh x_i} \prod_{i < j} \tanh\left({x_i-x_j\over 2}\right)^2, \quad {  g_s^{-1}={2  \Lambda \over \pi^{2}\epsilon} >0}~.\ee
Likewise the counterpart of  \eqref{zj} in this limit is
\be \label{zj4d}  \ba Z_2^{\rm 4d}(M)=&\ri^{-1} \int_{ \IR+\sigma_0}{\rd \sigma}  \tan \left(2 \pi \sigma\right) Z^{\rm Nek}(\sigma,t)\re^{- \log \left[2 \cos (2 \pi  \sigma )\right] M}  \re^{\frac{\log (2)}{12} +3\zeta'(-1)}t^{-1/16}\re^{4 \sqrt{t}} , \\
& \qquad \sigma_0=(2 \pi )^{-1}{\ri \cosh ^{-1}(2 \pi )} , \quad  t= \left(\frac{\Lambda }{4 \pi ^2 \epsilon }\right)^4,\ea\ee
Since $Z^{\rm Nek}(\sigma, t)$ has poles only for $\sigma \in \IZ/2$ one can take a generic $\sigma_0 \in \ri \IR_+$. 
This equality also follows from \eqref{4dstat} and \eqref{dic4}.  Indeed one has
\be \label{interm}Z_2^{\rm 4d}(M)={1\over 2 \pi \ri } \oint_0\kappa^{-N-1}\det (1+\kappa \rho^{\rm 4D}_{\IP^1\times \IP^1}) \rd \kappa.\ee
By using the change of variable \eqref{dic4} together with the identity  \eqref{4dstat}  we can write \eqref{interm} as \eqref{zj4d}.
 Moreover \eqref{zj4d} can also be tested numerically in an easy way thanks to the good convergent properties of \eqref{nek4d}. For instance on the l.h.s. we have
\be \label{tt}\ba& Z_2^{\rm 4d}(0)=1, \quad Z_2^{\rm 4d}(1)=\left(2 \pi \right)^{-1}{K_0\left(8 \sqrt[4]{t}\right)}, \\
& Z_2^{\rm 4d}(2)=\frac{G_{1,3}^{3,0}\left(64 \sqrt{t}|
\begin{array}{c}
 \frac{3}{2} \\
 0,0,0 \\
\end{array}
\right)}{32 \pi ^{3/2}},\ea\ee
where $K_0$ denotes the Bessel function and $G$ the Meijer G function. It is easy to check that the numerical integration on the r.h.s of \eqref{zj4d} reproduces \eqref{tt}. 

Equation \eqref{zj4d} provides the integral kernel implementing the exact S-duality transformation of pure $SU(2)$ Nekrasov partition in a self-dual $\Omega$-background parametrized by $\epsilon$. It is easy to see that in the semiclassical limit  $\epsilon\to 0$ it reduces to a simple Fourier transform as it is expected from Seiberg-Witten special geometry relations  \cite{sw2,abk}. Indeed, by introducing the SW variables $   {\tt a}/\epsilon\equiv \sigma $ and $ {\tt a}_D/\Lambda \propto g_s M$  one has in the $\epsilon\to 0$ limit 
\be 
\log 2 \cos \frac{2\pi {\tt a}}{\epsilon} =- \frac{2\pi \ri {\tt a}}{\epsilon}+ \mathcal{O}(\re^{4 \pi \ri {\tt a} \over \epsilon  }) \  \  , \quad \  \ 
  \ri^{-1}\tan \frac{2\pi {\tt a}}{\epsilon} =1+  \mathcal{O}(\re^{4 \pi \ri {\tt a} \over \epsilon  }) \ \ .
\ee
 Because of the imaginary part in the integration contour we have  ${\rm Re}(4 \pi \ri {\tt a} ) <0$, hence the corrections $\mathcal{O}(\re^{4 \pi \ri {\tt a} \over \epsilon })$ are exponentially suppressed and do not appear in the perturbative limit $\epsilon \to 0$. Notice also that
equation \eqref{zj4d}   agrees with the general philosophy for the change of frame in topological strings  \cite{abk} and provides the exact S-duality transformation properties of the gauge theory
partition function with gravitational corrections.

All this procedure can be in principle extended to $SU(2)$ theories with matter multiplets as well, however the details still have to be worked out.
Summarizing the implementation of the dual limit on the TS/ST duality leads to  the following  results in connection with four dimensional $\CN=2$ gauge theories: it gives an operator theory interpretation of the self-dual $\Omega$ background, it gives Fredholm determinant representation for the $\tau$ functions of  Painlev\'e equations, it provides a matrix model  for the partition function in the magnetic frame
and an exact formula for its S-duality transformation.

\section{Non-perturbative string on $Y^{N,0}$  geometries } \label{nps}

The TS/ST duality \cite{ghm} has been generalized to higher genus mirror curve in \cite{cgm2}.   
According to this construction one can associate a set of  $g$ operators \be \left\{{\rm O}_i \right\}_{i=1}^g \ee to any toric CY manifold, $g$ being the genus of its mirror curve.
Of particular interest for this paper are those CYs from which one can engineer  $SU(N)$ supersymmetric  gauge theories \cite{kkv,ikp3,ta}.  
Examples of such  geometries are  the resolution of the cone over the $Y^{N,0}$ singularity studied  for instance in  \cite{bz}.
  The corresponding mirror curve has genus $N-1$ and therefore  there are $N-1$ different "canonical" forms for this curve which reads 
\be O_i(x_1,x_2, \xi)+\kappa_i=0, \quad i=1, \cdots , N-1,\ee
where $\kappa_i$ denote the complex moduli of the geometry.
For instance we have
\be \label{cl1}O_1(x_1,x_2, \xi)+\kappa_1=\re^{x_2}+\re^{-x_2 +(-N+2)x_1}+\sum_{i=1}^{N-1} \kappa_{N-i} \re^{(i-N+1)x_1} +\xi \re^{(-N+1)x_1} +\re^{ x_1}=0 ,\ee  
where $\xi$ is the mass parameter and should be distinguished from the others moduli $\kappa_i$ as  emphasized for instance in  \cite{hkp}.
Therefore, the quantization procedure for  the $Y^{N,0}$ geometry  leads to the following $N-1$ operators 
\be \ba 
&{ {\rm O_1}+\kappa_1=\re^{\mathsf x_2}+\re^{-\mathsf x_2 +(-N+2)\mathsf x_1}+\sum_{i=1}^{N-1} \kappa_{N-i} \re^{(i-N+1) \mathsf x_1} +\xi \re^{(-N+1) \mathsf x_1} +\re^{\mathsf  x_1}},\\
&  {\rm O}_{j}+\kappa_{j}= {\rm Q}_j^{-1/2}\left( {\rm O}_1+\kappa_{1}\right) {\rm Q}_j^{-1/2}, \quad 1 < j \leq  N-1,
\ea\ee
where $[\mathsf x_1, \mathsf x_2]=\ri \hbar$
and we denote
\be {\rm Q}_j=\re^{ -(j-1) \mathsf x_1}.\ee
Let us define the following two operators 
\be  \label{rad}\ba&  \rho_{1, N-2,\xi}=\left(\re^{\mathsf x_2}+\re^{-\mathsf x_2 +(-N+2)\mathsf x_1}+\xi \re^{(-N+1)\mathsf x_1} +\re^{\mathsf  x_1}\right)^{-1}, \\
& {\rm A}_j^{\rm 5D}=\rho_{1, N-2,\xi}{\rm Q}_j.\ea\ee 
The conjecture \cite{cgm2} states that the non--perturbative   topological string partition function on  the $ Y^{N,0}$ geometry in the conifold frame is given by
\be
\label{gen-fred-th}
Z_N({M_1}, \cdots M_{N-1})= {1\over M_1! \cdots M_{N-1}!} \int  {\rm det}_{m,n} \left(R(x_m, x_n)\right)\rd^M x, 
\ee
where
\be
M=\sum_{j=1}^{N-1} M_j,
\ee
and 
\be
R (x_m, x_n) = A_{j}^{\rm 5D}(x_m , x_n) \,\, \,\, \,\, \,\,  \text{if} \, \,  \,\, \,\, \,\,\sum_{s=0}^{j-1} M_s\leq m \le \sum_{s=1}^j M_s.
\ee We use $M_0=1$ and 
\be A_{j}^{\rm 5D}(x_m , x_n)\ee
denotes the kernel of the operator  ${\rm A}_{j}^{\rm 5D}$ defined in \eqref{rad}.
The explicit expression for a  kernel of  the form \be\rho_{n,m,\xi }= \left(\re^{\mathsf  x_1}+ \re^{\mathsf  x_2} + \re^{-m \mathsf  x_1-n \mathsf  x_2} + \xi \re^{-(1+m)\mathsf  x_1-(n-1)\mathsf  x_2} \right)^{-1}\ee
was computed in \cite{cgum}. Let us review how this goes. We introduce some new variables $\mathsf q, \mathsf p$ such that \cite{kama,cgum} 
 \be \label{newv} \ba  \mathsf x_1= {2\pi b \over m+n+1}\left({\mathsf p+n \mathsf q }\right)\quad& \mathsf x_2=-{2\pi b \over m+n+1}\left({- \mathsf p+(m+1)\mathsf q }\right),   \ea\ee 
 where \be \hbar={2\pi \over m+n+1} b^2.\ee
 In particular we have
\be [\mathsf q, \mathsf p]={\ri \over 2 \pi}.\ee
 Then, the kernel of $\rho_{n,m,\xi }$ in the momentum representation w.r.t.~the new variables 
  reads \cite{cgum} 
\be  \label{rhomon} \rho_{n,m, \xi}(p, p')= {{\overline  f_\zeta(  p)}  f_\zeta(p')  \over 2 b \cosh\left( \pi {p-p'+\im h \over b} \right)},\ee
where
 \be
\label{fx}\ba 
f_\zeta(x)=&{\fad(x-\zeta + \ri n { c} ) \over \fad(x-\im (\alpha+c))} \re^{ 2 \pi (\alpha+ n c) x } \re^{ -2 \pi  c n  \zeta}, 
\ea\ee
 \be
\label{fx}\ba 
\overline f_\zeta(x)= &{\fad(x+\im (\alpha+c)) \over \fad(x-\zeta - \ri n c ) } \re^{ 2 \pi (\alpha+ n c) x } \re^{ -2 \pi  c n  \zeta}. \ea\ee
We  denote by $\fad $ the Faddeev's quantum dilogarithm \cite{faddeev,fk} and we use
\be
{  \alpha={b m\over 2(m+n+1)}}, \qquad c= {b \over 2(m+n+1)}, \qquad h = \alpha+c-nc, \qquad  \zeta={1\over 2 \pi b}\log \xi.
\ee
Some useful properties of the quantum dilogarithm can be found in Appendix A of \cite{kama}.
In our case we specialize the  above formulae to $n=1$ and $m=N-2$. Therefore, in the particular case of the $Y^{N,0}$ geometries, the partition function  \eqref{gen-fred-th} reads 
\be \label{gmat} \ba Z_N(M_1, \cdots,  M_{N-1})=&{1\over M_1! \cdots M_{N-1}! } \sum_{\sigma \in S_{M}}(-1)^\sigma\int  \rd ^M x \left(\prod_{i=1}^{M_1} A_1^{\rm  5D}(x_{\sigma(i)},x_i) \right)\\
&\left(\prod_{i=1+M_1}^{M_1+M_2} A_2^{\rm    5D}(x_{\sigma(i)},x_i)\right) \cdots
  \left( \prod_{i=1+\cdots +M_{N-2}}^{M_1+\cdots +M_{N-1}} A_{N-1}^{\rm   5D}(x_{\sigma(i)},x_i)\right)\\
\ea\ee
where 
\be \ba
 &A_j^{\rm   5D}(p,p')=\re^{-\ri  \pi b^2( j-1)^2/N^2}\re^{-2\pi ( j-1)b p'/N } \rho_{1, N-2,\xi}(p,p'+{\ri b ( j-1) \over N}), \\
\ea\ee
and \be  \rho_{1, N-2,\xi}(p,p') \ee is given by \eqref{rhomon}.
By using the Cauchy identity
\be \ba &{\prod_{1\leq i<j\leq N}2\sinh\left({\mu_i-\mu_j\over 2} \right)2\sinh\left({ \nu_i-\nu_j \over 2}\right) \over \prod_{i,j=1}^N2\cosh\left( {\mu_i-\nu_j \over 2}  \right)}= \sum_{\sigma\in S_N}(-1)^\sigma \prod_{i=1}^N{1\over 2 \cosh \left({\mu_i- \nu_{\sigma(i)}\over 2}\right)},\ea \ee
 together with some algebraic manipulations, we can write  \eqref{gmat}  as 
\be \label{mmsu35d}\ba Z_N(M_1, \cdots,  M_{N-1})=& { 1\over M_1!  \cdots M_{N-1}!} \int {{\rm d} ^M x\over (2\pi)^M} \prod_{j=1}^{N-1} \prod_{i_j\in I_j}\re^{V_j^{\rm 5D}(x_{i_j})} \\
 & \times {\prod_{1 \leq i<j \leq M}2\sinh\left({x_i-x_j\over 2}+{1\over 2} (d_i-d_j) \right)2\sinh\left({ x_i-x_j \over 2}+{1\over 2}(f_i-f_j)\right) \over \prod_{i,j=1}^M2\cosh\left( {x_i-x_j \over 2} + {1\over 2}\left( d_i-f_j\right) \right)}.
\ea \ee
We denote 
 \be \label{genn} \ba d_j = &-  {(N-1-k) \ri \pi  \over N}, \quad  \text{if}\quad  \sum_{s=0}^{k-1}M_s \leq j \leq  \sum_{s=1}^{k}M_s,\\
  f_j=&- {(N-2) \ri \pi  \over N} -d_j,
  \ea\ee
  and
\be \ba
\re^{V_k^{\rm 5D}(x)}= &\re^{- b^2  (k-1)x \over N}\re^{-{(k-1)\over 2N}\log \xi } {f_\zeta({b\over 2 \pi } x+\ri b  {(k-1)\over 2 N}+\zeta/2)} \overline{f}_\zeta({b\over 2 \pi } x-\ri b  {(k-1)\over 2 N}+\zeta/2). \ea\ee
Moreover
\be \label{inter} I_j=\left[{\sum_{s=0}^{j-1} M_{s}},\sum_{s=1}^{j} M_{s}  \right] \cap \IN , \quad M_0=1, \quad M=\sum_{i=1}^{N-1}M_i . \ee
According to  the conjecture \cite{ghm,cgm2,mz} the above matrix model  computes the topological string partition function of the $Y^{N,0}$ geometry in the conifold frame.  When $N=2$ this conjecture was tested in details in \cite{ghm,kmz} while for $N=3$ several tests have been performed in \cite{cgum}.  In the following we will compute the dual 4d limit of these matrix models for genetic $N$ and test the conjecture  by cross--checking  our result with some existing results in the gauge theory literature in particular with the works of  \cite{klemmlt,dp1,Edelstein:1999fz,Edelstein:2000aj}.

\section{ Four dimensional $SU(N)$ gauge theory} \label{sun4}
As described in section \ref{su2rev}, given the non-perturbative  topological string formulation of \cite{ghm, mz, cgm2},
there are two four dimensional limits in which we make contact with four dimensional $\CN=2$ gauge theory. 
In the {\it standard} four dimensional limit of the  $Y^{N,0}$ geometry  
 one takes  $\hbar \to 0$
 and it was shown in \cite{hm} that in this limit the TS/ST duality reduces to the well known four dimensional NS conjecture relating the spectral properties of $SU(N)$ quantum Toda to $\CN=2$ $SU(N)$ SYM in the NS phase of the $\Omega$ background \cite{ns}.   
In the {\it dual} four dimensional limit instead, we take
\be \label{dlg} \ba & \hbar={  1 \over \epsilon  \beta}, \quad \log \xi ={\hbar \over 2\pi } \left(a \epsilon \beta -\log(\beta^{2N}\Lambda^{2N}) \right),  \\
 &\log \kappa_i=- {\hbar \over 2 \pi N} \log \left(\beta^{2N} \Lambda^{2N}\right)+\log  \left(G_i\right) + \mathcal{O}(\re^{\beta^{-1}\log \beta^4}), \quad \beta \to 0^+ . \ea
\ee
After implementing this limit we make contact with $d=4$, $\CN=2$, $SU(N)$ SYM in the self--dual  $\Omega$ background. The parameter $\epsilon$ is the twisting parameter of the $\Omega$ background,  $\Lambda$ is the instanton counting parameter while $a, G_i$ are related to the A  periods of the SW curve underling the  four dimensional theory.

\subsection{The matrix model }

In this section we implement the dual limit \eqref{dlg} on the matrix model  \eqref{mmsu35d}.
We obtain
\be \left(\prod_{i=1}^{N-1}\kappa_i^{M_i}\right)Z_N(M_1, \cdots,  M_{N-1})\xrightarrow{\text{ dual 4d} }\left(\prod_{i=1}^{N-1}\left(\re^{-a/( 2N \pi)}G_i\right)^{M_i}\right) Z_{\rm N}^{\rm 4d}(M_1,\cdots,M_{N-1})\ee
where 
\be \label{su3mm} {\ba Z_{\rm N}^{\rm 4d}(M_1,\cdots,M_{N-1})=&  {1 \over M_1!  \cdots M_{N-1}!} \int {{\rm d} ^M x\over (2\pi)^M} \prod_{j=1}^{N-1} \prod_{i_j\in I_j}\re^{{ -{1\over g_s}}V_j (x_{i_j})}  \\
&   \times {\prod_{1\leq i<j \leq M}2\sinh\left({x_i-x_j\over 2}+{1\over 2} (d_i-d_j) \right)2\sinh\left({ x_i-x_j \over 2}+{1\over 2}(f_i-f_j)\right) \over \prod_{i,j=1}^M2\cosh\left( {x_i-x_j \over 2} +{1\over 2} (d_i-f_j), \right)}.
\ea }\ee
The quantities $f_i, d_i$  are as in  \eqref{mmsu35d}. Likewise $I_j$ is defined in \eqref{inter} while the potential  is now given by
\be \ba V_{k}(x)=& { \sin \left(\frac{\pi  k }{N}\right)\over  \sin \left(\frac{\pi  }{N}\right)} \cosh(x) ,\ea\ee 
and \be \label{gsdef}  g_s = \frac{ \pi ^2 \epsilon}{N \Lambda  \sin \left(\frac{\pi   }{N}\right) }.
\ee
 Hence in the dual 4d limit we recover a $(N-1)$-cut matrix model, whose   't Hooft couplings are
\be  T_i= { g_s M_i }.\ee
According to \cite{bgt} to make contact with gauge theory we identify the dual periods $a_D^{(i)} $ with the 't Hooft couplings, namely  
\be a_D^{(i)} = T_i .\ee
By combining the   TS/ST conjecture \cite{ghm,cgm2,mz}  with the work of \cite{bgt} we expect that the matrix model  \eqref{su3mm}  computes the partition function of $d=4$, $SU(N)$ $\CN=2$ SYM  in the self--dual $\Omega$ background  in the magnetic frame.  Therefore from our perspective the rank of the gauge group is related to the number of cuts of the matrix model and not to the sizes of the matrices.
  The above conjecture can be tested  against existing results in the literature.
In particular we will test the 't Hooft expansion of \eqref{su3mm} for small values of the 't Hooft couplings. This expansion is obtained by considering \be M_i, g_s^{-1} \to \infty, \quad T_i =M_i g_s \quad \text{fixed}.\ee
In this limit the matrix model \eqref{su3mm} displays the following perturbative behavior
\be \label{thooft1} \log Z_N^{\rm 4d}(M_1, \dots, M_{N-1} )={  \sum_{g\geq 0}g_s^{2g-2}F_g^{ D}(T_1,\cdots,T_{N-1})}, \ee
where $ F^D_g (T_1,\cdots,T_{N-1})$ are the genus $g$ free energies.
On the other hand, as explained for instance in  \cite{akmv,gm,cgum}, we can expand  the matrix model around the gaussian point for small values of the $g_s$ and fixed values of $M_i$. One finds  that the   logarithm of the partition function has the following behavior
\be\label{weak}\ba &  \log Z_N^{\rm 4d}(M_1, \dots, M_{N-1} )=\sum_{i=1}^{N-1}{1\over 2}\left( M_i^2\log \left(g_s M_i\right)\right)+
\mu_i \log M_i+ \zeta\\
&+\sum_{n\geq -1}g_s^{n}\sum_{i_1,\cdots, i_{N-1}}' \CF_{(n-i_1-\cdots-i_{N-1})/2+1,i_1,\cdots,i_{N-1}}M_1^{i_1}\cdots M_{N-1}^{i_{N-1}},
\ea\ee
where $\sum_{i_1,\cdots, i_{N-1}}'$ runs only over $i_1,\cdots, i_{N-1}$ such that
\be\ba {i_1,\cdots, i_{N-1}}\geq {\bf {0}}, & \quad  \left\{i_1,\cdots, i_{N-1}\right\}\neq {\bf {0}}, \quad 
\sum_k i_k=n \mod 2,
\ea\ee
while $\mu_i, \zeta$ are constants. By performing the gaussian integration for severals fixed values of $M_i$, the coefficients $\CF$ appearing in \eqref{weak} can be computed exactly.
By combining the two expansions \eqref{thooft1}, \eqref{weak} it follows that the small $T_i$ expansion of the genus $g$ free energy
 \be F_g^D (T_1,\cdots,T_{N-1})\ee is determined by  $\CF_{g,i_1,\cdots,i_{N-1}}$. 
For instance we have  (see appendix \ref{testsu3} for more details)
\be \label{fogen} \ba F_0^D (T_1, \cdots T_{N-1})=&\sum_{i=1}^{N-1} \left( {1\over 2}{T_i}^2\log T_i \right) +\sum_{ i_1, \cdots  i_{N-1}}'\CF_{0, i_1, \cdots, i_{N-1}}T_{1}^{i_1}\cdots T_{N-1}^{i_{N-1}}.
\ea\ee
Let us start by applying  these techniques  to compute the genus zero free energy \eqref{fogen} of the matrix model  \eqref{su3mm}.  We obtain 
\be \label{myf0n} \ba F_0^D(T_1,\cdots, T_{N-1})=&\sum_{j=1}^{N-1}\left[{T_j^2\over 2}\left(\log \left( {T_j \over c_j}\right) -{3\over 2}\right) - T_j {  \sin\left({\pi j \over N}\right) \over   \sin\left({\pi  \over N}\right)}\right]+\sum_{i<j}{  T_i T_j \log \left(a_{ij}\right)}  + \mathcal{O}(T_i^3),\ea \ee
where
\be \label{coefac}\ba a_{ij}=&\frac{\cos \left(\frac{\pi  (i-j)}{N}\right)-1}{\cos \left(\frac{\pi  (i+j)}{N}\right)-1}, \\
c_j=&4 \sin\left(\frac{\pi  j}{N}\right)^3 /\sin \left(\frac{\pi }{N}\right), \ea\ee
and $\mathcal{O}(T_i^3)$ are the so called instantons  corrections to the prepotential and are computed  in appendix \ref{testsu3}. 
The result in \eqref{myf0n} and \eqref{coefac} agrees with the computation of   the $SU(N)$  $\CN=2$ prepotential  obtained in \cite{klemmlt,dp1, Edelstein:1999fz,Edelstein:2000aj}
\footnote{  The agreement is up to the the  $c_j$ term. Our normalization agrees with the one in \cite{klemmlt}. }.
In particular the off diagonal terms of  the prepotential namely 
\be \sum_{i<j}{  T_i T_j \log \left(a_{ij}\right)} \ee
have been computed for $SU(N)$ gauge theories in \cite{Edelstein:1999fz, Edelstein:2000aj}. We checked that our result \eqref{coefac} agrees with the one of \cite{Edelstein:1999fz,Edelstein:2000aj} namely
\be \log \left(a_{ij}\right)=2 { \pi \ri } \tau_{\rm D}^{ij},\ee
where $\tau_{\rm D}^{ij}$ is defined in \cite{Edelstein:2000aj}.

Let us now look at the instanton part of the prepotential. 
The case $SU(2)$ was discussed in details in \cite{bgt} where it was shown that the matrix model \eqref{su3mm} matches  the existing results in the gauge theory literature which include higher genus $F_g$ computations as well.
For $SU(3)$  several computations have been performed in the 90's on the gauge theory side in particular in \cite{klemmlt}. As explained  in appendix \ref{testsu3}  the matrix model \eqref{su3mm} reproduces all these results. 
For generic $SU(N)$ theories the first instanton correction to \eqref{myf0n} was computed in \cite{dp1}.  We  have checked explicitly  that  the matrix model \eqref{su3mm} reproduces  the results of  \cite{dp1} if $N=4$ and $N=5$. 

As an additional observation we briefly comment on other proposals for matrix models describing  $SU(N)$ SYM theories. 
Let us first underline that all of them aim to describe the gauge theory partition function in the weak coupling {\it electric} frame while the one proposed here computes the gauge theory partition function in the strongly coupled {\it magnetic} frame. 

The idea of computing partition functions of supersymmetric gauge theories as matrix models goes back to the work of 
 Dijkgraaf-Vafa \cite{dv}. Even though the original formulation applies  to $\CN=1$ theories, one can obtain some results in $\CN=2$ theories  as well  (see for instance \cite{cv}). In the case of  $d=4$, $\CN=2$, $SU(N)$ theories  it was found in  \cite{kmt} that 
 these matrix models are expected 
 to reproduce  only the first few terms in the perturbative expansion of the gauge theory, namely  only $F_0$ and $F_1$. 
 Our models instead reproduce the full dual free energies $F^D_g$s of the gauge theory
in the self-dual $\Omega$ background. For instance in the $SU(2)$ case we have checked explicitly that the genus $g$ free energies $F^D_g$ of \eqref{su3mm} agree with those of the gauge theory (computed in \cite{hk06})  up to genus  $g=4$. 

 On the other hand a different class of models was also proposed for instance in \cite{ksp} \footnote{ A more exhaustive list of references as well as a detailed discussion on the limits of these other models can be found in \cite{mz}.}. These proposals are very different from our one and are constructed by elaborating from the Nekrasov partition function itself. This  leads to  formal models which are  not well defined beyond perturbation theory.  Our matrix models instead, as well as the ones in \cite{mz,kmz,cgm2,bgt},  are obtained by quantizing mirror curves and in particular without using  Nekrasov results. Therefore the fact that we reproduce the gauge theory expansion  is highly--non trivial. Moreover our models are  well-defined matrix models also beyond perturbation theory i.e.~at finite $M$ and finite $g_s$ as it is manifest from \eqref{su3mm} and \eqref{mmsu35d}. 

\subsection{Operator theory }\label{opno}
In this subsection  we give an explicit expression for the operators behind the dual four dimensional limit of the the $Y^{N,0}$ geometry.
By  using the Cauchy identity 
we can rewrite  the matrix model \eqref{su3mm}  as 
\be \label{gmat4d} \ba  Z^{\rm 4d}_N(M_1, \cdots,  M_{N-1})=&{1\over M_1! \cdots M_{N-1}! } \sum_{\sigma \in S_{M}}(-1)^\sigma\int  \rd ^M x \left(\prod_{i=1}^{M_1} A_1(x_i,x_{\sigma(i)}) \right)\\
&\left(\prod_{i=1+M_1}^{M_1+M_2} A_2(x_i,x_{\sigma(i)})\right) \cdots
  \left( \prod_{i=1+\cdots +M_{N-2}}^{M_1+\cdots +M_{N-1}} A_{N-1}(x_i,x_{\sigma(i)})\right)
\ea\ee
where $M=\sum_{i=1}^{N-1}M_i$ and  
\be A_{j}(x,y)= {     f_{\rm 4d}(x+\frac{\ri \pi  (4j-N-2)}{2 N})  f_{\rm 4d}(y+\ri\pi   {N-2\over2 N})  \over 4 \pi \cosh \left({x- y\over 2}+\frac{i \pi  (2 j-N)}{2 N}\right)}, \quad j=1, \cdots, { N-1}
 \ee with
\be  \ba
 f_{\rm 4d} (u) = \exp \left(-\frac{1}{ 2 g_s  \sin \left(\frac{\pi   }{N}\right) } \cosh (u)\right) . \\
%\tilde f_{\rm 4d}=
\ea \ee
These are the kernels (in the momentum representation) of the following operators 
\be \ba    A_{1}&=  \re^{ - {N-2\over 4 N}{ \mathsf q  } }{ f_{\rm 4d}(\mathsf p)}{ 1 \over 2 \cosh ( \mathsf q/2) } {  f_{\rm 4d}(\mathsf p)}  \re^{ - {N-2\over 4 N}{ \mathsf q  } },  \\
 A_{j}&=  P_{j} A_1, \quad P_j=\re^{  {(j-1)\over  N} {\mathsf q} }, \qquad j=1, \cdots, N-1, \qquad [\mathsf q, \mathsf p]= 2 \pi \ri
\ea \ee
Hence
\be \label{det4}\sum_{{ {M_i}}\geq 0} x_1^{M_1} \cdots x_{N-1}^{M_{N-1}}  Z^{\rm 4d}_N(M_1, \cdots,  M_{N-1})=\det \left( 1+x_1 A_1+\cdots+ x_{N-1}A_{N-1} \right) \ee
where, in terms of the dictionary \eqref{dlg}, we have \be x_i = \re^{-a/( 2N \pi)}G_i.\ee
Notice that the operators $A_j$ are of trace class hence the determinant  \eqref{det4} is well defined and analytic in the $x_i$'s.
Furthermore  by combining  \cite{ghm,cgm2,bgt}  with the above computation, it follows that the  Nekrasov--Okounkov \cite{no2} partition function $ Z^{\rm NO}$ for pure  $d=4$, $\CN=2$, $SU(N)$ SYM  in the self--dual background has a spectral determinant interpretation, namely\footnote{  In this identification one has to be careful in imposing a suitable  domain of definition on the parameters $ x_i$  in the l.h.s. on the equality. For instance in the $SU(2)$ case we usually consider $\sigma \neq \IZ/2$ otherwise an appropriate regularisation of $Z^{\rm NO}$ could be required (see \cite{bgt}). }
\be  \label{4dsd}\ba Z^{\rm NO}(x_1,\cdots, x_{N-1}, \Lambda, \epsilon)=\CN(\Lambda, \epsilon)\det \left( 1+x_1 A_1+\cdots+ x_{N-1}A_{N-1} \right), \ea\ee
where $\CN (\Lambda, \epsilon)$ is a non-vanishing proportionality constant.  From the topological string viewpoint this constant is related to the constant map contribution as well as the classical polynomial part of the topological string free energy.
For instance in the case of $SU(2)$ we have \cite{bgt}
 \be \CN (\Lambda, \epsilon)=\re^{- \Lambda^2/(4 \pi^4 \epsilon^2)} \left({\Lambda\over 4 \pi^2 \epsilon}\right)^{1/4} \re^{-{1\over 12}\log 2-3 \zeta' (-1)}.\ee
Hence the spectrum of the above operators is completely determined by  the $\CN=2$, $SU(N)$ SYM in the self--dual $\Omega$ background. 
 Moreover, as discussed in section \ref{newsection} the spectral determinant representations of the NO partition function determines its exact S-duality transformation. In the $SU(2)$ case this is given by \eqref{zj4d} and  a similar behaviour also holds in the $SU(N)$ case. Indeed we have
\be \label{zdet4}Z^{\rm 4d}_N(M_1, \cdots,  M_{N-1})= {1 \over (2 \pi \ri)^{N-1}}\oint _0 {\rd x_1 \over x_1^{M_1+1}}\cdots \oint _0 {\rd x_{N-1} \over x_{N-1}^{M_{N-1}+1}}\det \left( 1+x_1 A_1+\cdots+ x_{N-1}A_{N-1} \right).\ee
By combining  \eqref{zdet4} with \eqref{4dsd} one obtains  the exact S-duality transformation which generalises   \eqref{zj4d} in the higher rank case.
Notice that, after a suitable parametrization, the classical version of \be 1+\sum_{i=1}^{N-1}x_i A_i=0\ee
coincides with the classical $SU(N)$ Seiberg-Witten curve, namely \be \label{si} z+z^{-1}-v^N+\sum_{i=0}^{N-2}x_{N-i-1} v^i=0,\ee
where we used \be z+z^{-1}=2\cosh(q/2) \exp \left[{-{ N-2\over 2 N}q+{N\Lambda\over \pi^2 \epsilon} \cosh p}\right] +\re^{-q}, \quad v=\re^{-q/N}. \ee

As an additional observation we  recall that in the pure $SU(2)$ case the  determinant \eqref{4dsd} corresponds to the $\tau$ function of  the Painlev\'e $\rm III_3$ \cite{ilt,bgt,zamo}. Moreover, by changing the matter content of the $SU(2)$ theory, we  recover the $\tau$ function for  the other Painlev\'e equations  \cite{blmst}.   It is therefore natural to expect that in the generic $SU(N)$ case  \eqref{4dsd} provides the solution for an isomonodromy problem which generalizes the one related to Painlev\'e equations. 

According to \cite{blmst} the connection of which one has to study the isomonodromy problem is obtained from the Hitchin system describing the relevant Seiberg-Witten geometry. 
In the forthcoming subsection \ref{opp} we will describe the isomonodromy problem associated to the one-period phase
of the matrix model \eqref{su3mm} which as we will see give rise to the closed Toda chain equations. We remark that in order to get the appropriate connection one has to 
consider Hitchin's system on a cylinder: this is in line with the findings of 4d/2d correspondence discussed in \cite{cv} which naturally connects the isomonodromy problem to $tt^*$ equations \cite{ttstar}.

\subsection{The  one period phase }
\label{opp}

Let us consider the particular point in the moduli space of the $SU(N)$ theory where only one dual period is non vanishing.
We can think of this point as a 1-period deformation of the  strong coupling singularity described in \cite{Douglas:1995nw}.
  At this singularity the $N-1$ monopoles become massless and the dual periods of the $SU(N)$ Seiberg--Witten curve vanish
   \be a_D^{(k)}=0, \quad k=1, \cdots N-1.\ee  
If we keep only one non--vanishing period, namely
 \be a_D^{(1)} \neq 0, \ee the partition function is described by the following one-cut matrix model 
 \be \label{ad0}\ba Z_{\rm 4d}^{(1)}(M)= &{ 1 \over M! }  \int {{\rm d} ^M x\over (2\pi)^M} \prod_{i=1}^{M } \re^{ -\frac{1}{ g_s }  \cosh (x_i)} {\prod_{i<j}4\sinh\left({x_i-x_j\over 2}  \right)^2\over \prod_{i, j}2\cosh\left( {x_i-x_j \over 2} +\ri \pi \beta, \right)}, \quad \beta={2-N\over 2 N}.\ea \ee
This matrix model computes  the fermionic spectral traces of the following 
kernel 
\be\label{k1} { \rm K}(x,y)={\re^{-\cosh (x)  /2 g_s}}{ 1 \over 4 \pi \cosh \left( {x-y\over 2}- \ri \pi {(N-2) \over 2N}\right) } {\re^{-\cosh(y)  /2 g_s}}, \quad x,y \in \IR,\ee
namely
\be Z_{\rm 4d}^{(1)}(M)= \sum_{\sigma\in S_M}(-1)^{\sigma} { 1 \over M! }  \int {{\rm d} ^M x}   \prod_{i=1}^{M }  K(x_i, x_{\sigma(i)}). \ee
Hence, it follows from \cite{wid} that 
 \be \label{qdef}q_k=\log \left[\det(1-\kappa \re^{2(k+1)\ri \pi/N }K)/\det(1-\kappa \re^{2k\ri \pi/N} K)\right], \quad k \in \IZ\ee
 fulfills the first equation of Toda hierarchy which reads
 \be \label{qt} q_{\ell} ^{''}+{1\over t}q_{\ell}^{'}= \re^{q_{\ell}-q_{{\ell}-1}}-\re^{q_{{\ell}+1}-q_{{\ell}}}, \ee
where  \be t=(2 \sin \left({\pi/N}\right)g_s)^{-1} .\ee
 It was found in  \cite{its1,its2} that \eqref{qt} arises as compatibility condition for the isomonodromy deformations of an $SL(N,\IC)$ connection on a cylinder with regular singularities at zero and at infinity. This agrees with our general expectation (see discussion at the end of the previous subsection) and it gives a concrete relation between the $SU(N)$ matrix models presented above and the tt* equations of \cite{ttstar}.
More precisely, by considering the $N$-covering $z=w^N$ of the SW geometry \eqref{si} 
%and rotating to the $J$-complex structure, 
one gets the 
radial component of the 
Hitchin's connection as (see Chapter 5 of \cite{swbook})
\be
\CA=\frac{\partial}{\partial r} {\bf q} + w^{-1}\left(e^{-\alpha_0 q}E_{-\alpha_0}+\sum_{\alpha\,\, {\rm simple}} e^{\alpha q} E_\alpha\right)
+ w\left(e^{-\alpha_0 q}E_{\alpha_0}+\sum_{\alpha\,\, {\rm simple}} e^{\alpha q} E_{-\alpha}\right)
\ee
where ${\bf q}={\rm diag}(q_0,\ldots,q_{N-1})$ is subject to the traceless condition $q_0+\ldots+q_{N-1}=0$ and $\alpha$'s are the roots of  $\hat A_{N-1}$  algebra, $\alpha_0$ being the one associated to the affinization. The above is precisely the connection whose isomonodromy problem gives rise 
to closed Toda chain equations as explained in \cite{its1,its2}. Notice that this is describing the $tt^*$ geometry of the  Landau-Ginzburg model  whose potential is associated to the $A_{N-1}$ singularity.
In the one-period case one considers isomonodromy with respect to the length of the cylinder which corresponds to the Yang-Mills coupling of gauge theory. We expect the higher times of Toda hierarchy
to be related with the insertion of local observables giving rise to new terms in the potential of the kind $T_n \cosh (n x)$, $T_n$ being related to higher times of Toda hierarchy.
On the other hand, the full isomonodromy problem considering Whitham deformations of the Hitchin's system \cite{jimbo2,Krichever:1992qe,Nakatsu:1995bz,Itoyama:1995nv, Gorsky:1995zq,Edelstein:1999xk,Bonelli:2009zp} 
should give rise to a system of PDEs satisfied by the gran-canonical partition function of the full matrix model \eqref{su3mm}.

In the particular case of $N=2$ the partition function \eqref{ad0} becomes the well studied polymer matrix model \cite{zamo,gm, fadsal,cfiv,ceva} and we have 
\be q_k=q_{k+2}=-q_{k+1} .\ee
As expected  \eqref{qt} becomes the $\sinh$ form of the Painlev\'e $\rm III_3$ equation
 \be \label{qt2} q_{0} ^{''}+{1\over t}q_{0}^{'}= \re^{2q_0}-\re^{-2 q_{0}}. \ee

Interestingly  this is not the first time than the matrix model \eqref{ad0}  appeared in the literature, indeed it is of the same type as the one considered by Kostov  in   \cite{kos} to describe the 6-vertex model. 
In \cite{oz} it was shown that such a matrix model can be evaluated exactly for fixed values of $M$ and $\hbar$  by using a TBA system which generalizes the one of  \cite{cfiv,zamo,tw}.
Moreover in \cite{Hollowood:2003gr} it was argued that the  matrix model \eqref{ad0} computes the partition function of $SU(M)$ $\CN=1^*$ SYM   in 5 dimensions. More precisely to relate \eqref{ad0} to  \cite{Hollowood:2003gr} we need the following dictionary  
 \be {2  \pi \over N}= R m,\ee
where in the context of  \cite{Hollowood:2003gr}  $R$ is the radius of the fifth dimension and $m$ the mass. To our knowledge this is a curious coincidence.  
 Moreover if we  rescale $$x \to 2 x \pi /N$$  and we take   $N \to \infty$ then \eqref{ad0} reduces to the  D-particle matrix model studied in  \cite{kkn} where  the  authors show that the corresponding spectral determinant is related to the $\tau$ function of the KP hierarchy.
We also observe that in  \cite{hkksym} a similar type of model, but with a slightly different contour,  was used in   to compute the partition function of the one dimensional reduction of $\CN=1$ $SU(M)$ SYM on a circle  of radius $\beta$ with insertion of a source term.
The same matrix model was also used in the context of 2d black hole in \cite{Kazakov:2000pm}. 
 It would be interesting to understand if this is just a mathematical coincidence or if there is a deeper physical meaning.

\section{Conclusions}
\raggedbottom
In this paper we proposed a matrix model computing  the partition function of $d=4$ $\CN=2$, $SU(N)$ SYM in the magnetic frame, coupled to self-dual $\Omega $ background. In particular the 't Hooft expansion of this matrix model reproduces the genus expansion of the gauge theory in the magnetic frame.   Our method relies on the formalism developed in \cite{cgm2}  for the quantization of  higher genus mirror curves and  as a consequence it provides a  spectral determinant representation for the  four dimensional Nekrasov--Okounkov partition function in the case of higher rank gauge theories, generalizing the results of \cite{bgt}.  Such a spectral determinant representation also  provides  exact S-duality transformation properties for the pure  gauge theory, which we have  spelled out in detail in the $SU(2)$ case, see \eqref{zj4d}. 

As explained in \cite{bgt}, when the gauge group is $SU(2)$ the matrix model  \eqref{su3mm} coincides with a well known $O(2)$ matrix model \cite{kos39,kos} which is related to the 2d Ising model and the physics of 2d self--avoiding polymers. It would be interesting to see if the  higher rank matrix model \eqref{su3mm}  can be connected to statistical systems too.

In the one-period phase, the spectral determinants \eqref{qdef} satisfy Toda lattice equation, see subsection \ref{opp}. We expect the corresponding hierarchy of differential equations to be satisfied by the spectral determinants associated to the generating function of the local chiral observables  (see for instance \cite{Fucito:2005wc}). It would be interesting to further explore this aspect.  Also it is worth investigating the insertion of surface operators (see for instance \cite{Ashok:2017odt}) and their description in the magnetic phase of the gauge theory as eigenfunctions of the quantum operators arising from the dual four-dimensional limit of the non-perturbative open topological strings \cite{Marino:2016rsq}. 

 Furthermore the matrix models proposed here are obtained by combining the quantization of mirror curves with the Fermi gas formalism \cite{mp,ghm,mz,cgm2,bgt}, however it would be important to derive them directly in gauge theory.

Another interesting point would be to prove explicitly  that the higher rank NO partition function is proportional to a suitably defined $\tau$ function for the isomonodromy problem related to the Hitchin's integrable system corresponding to  the $SU(N)$ SW geometry.  Finally, the extension of our results to gauge theories coupled to matter is still to be worked out. We hope to report on this in the near future. 

\section*{Acknowledgements}
We would like to thank Marco Bertola, Simone Giacomelli, Jie Gu, Rajesh Gupta, Victor I. Giraldo , Kohei Iwaki,  Vladimir Kazakov, Oleg Lisovyy and Marcos Mari\~no for valuable  discussions and in particular Jie Gu for a careful reading of the draft. 
A.G. thanks  Durham University, ENS Paris, Kyoto and Osaka Universities for the kind hospitality during the course of this project.
A.T. thanks IIP, Natal and Universit\'e Paris VII for kind hospitality during the course of this project.
The work of G.B. is supported by the PRIN project "Non-perturbative Aspects Of Gauge Theories And Strings".
G.B. and A.G. acknowledge support by INFN Iniziativa Specifica ST\&FI.
A.T. acknowledges support by INFN Iniziativa Specifica GAST.

%\newpage
\appendix

\section{More on the SU(3) and SU(4) matrix models } \label{testsu3}

In this appendix we provide  some details of computation  for the $SU(3)$ and $SU(4)$ cases. More precisely we test  by an explicit computation that the matrix model
\eqref{su3mm} for $N=3$ reproduces the prepotential  of $\CN=2$ $SU(3)$ SYM  as given in \cite{klemmlt}.  For $SU(4)$ we cross--check our results by comparing with  \cite{dp1}.

Let us start with the $SU(3)$ case. The matrix model \eqref{su3mm} is a 2-cuts  model which reads
\be \label{caso3}{\ba Z_{\rm 3}^{\rm 4d}(M_1,M_2)=&  {1 \over M_1!  M_{2}!} \int {{\rm d} ^M x\over (2\pi)^M} \prod_{i=1}^{M_1+M_2} \re^{-  \frac{3 \Lambda }{ \pi ^2 \epsilon}{ \sin \left(\frac{\pi   }{3}\right)} \cosh(x_{i})} \\
% \prod_{i_2=1+M_1}^{M_1+M_2}  \re^{-  \frac{3 \Lambda }{ \pi ^2 \epsilon}{ \sin \left(\frac{2\pi   }{3}\right)} \cosh(x_{i_2})}   \\
&   \times {\prod_{ 1\leq i<j \leq M}2\sinh\left({x_i-x_j\over 2}+{1\over 2} (d_i-d_j) \right)2\sinh\left({ x_i-x_j \over 2}+{1\over 2}(f_i-f_j)\right) \over \prod_{i,j=1}^M2\cosh\left( {x_i-x_j \over 2} +{1\over 2} (d_i-f_j) \right)},
\ea }\ee
where
\be  d_{i} =  \left\{\begin{array}{c c}
-{  \ri \pi  \over 3}   \quad \text{if}\quad i \leq M_1  , 
\\
0 \quad \text{if}\quad i > M_1   ,
 \end{array}\ \right.
 \qquad   f_{i} =  \left\{\begin{array}{c c}
0 \quad \text{if}\quad i \leq M_1   ,
\\
-{  \ri \pi  \over 3}   \quad \text{if}\quad i > M_1   .
 \end{array}\ \right.
\ee
By using standard techniques in matrix models (see for instance \cite{akmv,gm,cgum}) we can write the small $g_s$ expansion of a 2-cuts matrix model as  \eqref{weak}. In our case this becomes 
\be \label{zlog}\ba  & \log(Z_{\rm 3}^{\rm 4d}(N_1,N_2))=F^D(N_1,N_2)\\
&=  g_s^{-2}\left({1\over 2}T_1^2 \left( \log (T_1 /3) -{3 \over 2}\right)+{1\over 2}T_2^2 \left( \log (T_2 /3)-{3\over 2}\right)- (T_1+T_2)  -{T_1 T_2 \log 4 } \right)\\
&+ g_s\Big(N_1^3 \CF_{0,3,0}+N_1^2 N_2 \CF_{0,2,1}+N_1 N_2^2 \CF_{0,1,2}+N_1 \CF_{1,1,0}+N_2^3 \CF_{0,0,3}+N_2 \CF_{1,0,1}\Big)  \\
&+ g_s^2 \Big( N_1^4 \CF_{0,4,0}+N_1^3 N_2 \CF_{0,3,1}+N_1^2 N_2^2 \CF_{0,2,2}+N_1^2 \CF_{1,2,0}+N_1 N_2^3 \CF_{0,1,3}\\
&
+N_1 N_2 \CF_{1,1,1}+N_2^4 \CF_{0,0,4}+N_2^2 \CF_{1,0,2}+\CF_{2,0,0}\Big) \\
&+ g_s^3 \Big( \CF_{0,0,5} \left(N_1^5+N_2^5\right)+\CF_{0,1,4} \left(N_1^4 N_2+N_1 N_2^4\right)+\CF_{1,0,3} \left(N_1^3+N_2^3\right)+\CF_{1,1,2} \left(N_1^2 N_2+N_1 N_2^2\right) \\
&+\CF_{0,2,3} \left(N_1^3 N_2^2+N_1^2 N_2^3\right)+\CF_{2,0,1} (N_1+N_2)\Big) +\mathcal{O}( g_s^4),
\ea\ee
where $\CF_{g,i,j}$  determine the weak coupling expansion of  the free energy $F_g^D$ and we used
\be T_i=g_s N_i\ee
and $g_s$ is defined in \eqref{gsdef}.
 As it is manifest from  \eqref{caso3} the  matrix model  is symmetric under $T_1 \leftrightarrow T_2$. Then from \eqref{thooft1} and \eqref{weak} we have 
\be\label{fmm}\ba  F_0^D(T_1,T_2)= &{1\over 2}T_1^2 \left( \log (T_1 /3) -{3 \over 2}\right)+{1\over 2}T_2^2 \left( \log (T_2 /3)-{3\over 2}\right)- (T_1+T_2)  -{T_1 T_2 \log 4 } 
\\
&+\Big( \CF_{0,3,0} (T_1^3+T_2^3)+\CF_{0,1,2} (T_1^2 T_2+T_1 T_2^2) \Big)\\
&+ \Big( \CF_{0,0,4} \left(T_1^4+T_2^4\right)+\CF_{0,1,3} \left(T_1^3 T_2+T_1 T_2^3\right)  +\CF_{0,2,2}\left(T_1^2 T_2^2 \right)   \Big) \\
&+ \Big( \CF_{0,0,5} \left(T_1^5+T_2^5\right)+\CF_{0,1,4} \left(T_1^4 T_2+T_1 T_2^4\right)+\CF_{0,2,3} \left(T_1^3 T_2^2+T_1^2 T_2^3\right)\Big) 
+ \mathcal{O}{(T^6)}.\ea\ee
By computing  $Z(N_1,N_2)$  for various  values of $N_1, N_2$ we  can fix the  $\CF_{n,i,j}$ coefficients in \eqref{zlog}.
It follows that
\be \ba  F_0^D(T_1,T_2)= &{1\over 2}T_1^2 \left( \log (T_1 /3) -{3 \over 2}\right)+{1\over 2}T_2^2 \left( \log (T_2 /3)-{3\over 2}\right)- (T_1+T_2)  -{T_1 T_2 \log 4 } 
\\&  -{1\over 3} (T_1^3+T_2^3)+{3\over 4} (T_1^2 T_2+T_1 T_2^2) 
 \\
& +\frac{11}{36} \left(T_1^4+T_2^4\right)-\frac{23}{16} \left(T_1^3 T_2+T_1 T_2^3\right)-\frac{21}{16} T_1^2 T_2^2 \\
&-\frac{14}{27} \left(T_1^5+T_2^5\right)+\frac{731}{192} \left(T_1^4 T_2+T_1 T_2^4\right)+\frac{47}{16} \left(T_1^3 T_2^2+T_1^2 T_2^3\right) \\
&+ \frac{127 T_1^6}{108}-\frac{18299 T_1^5 T_2}{1536}-\frac{10105 T_1^4 T_2^2}{1536}-\frac{2777 T_1^3 T_2^3}{384}-\frac{10105 T_1^2 T_2^4}{1536}\\
&-\frac{18299 T_1 T_2^5}{1536}+\frac{127 T_2^6}{108}+\mathcal{O}{(T^7)}
\ea\ee
which matches the gauge theory result of \cite{klemmlt} once we identify the 't Hooft parameters of the matrix model with the dual periods.  Moreover, given the matrix model expression \eqref{caso3}, it is relatively easy to  compute $F_g^D$  for higher genus as well and, to our knowledge, these are not known in the gauge theory literature. For instance we have
\be  \ba F_1^D(T_1,T_2)=& -{1\over 12} \log \left(T_2T_1\right) + \frac{13 T_1 T_2}{16}+\frac{5 }{24}(T_1+T_2)-\frac{35}{144}  \left(T_1^2+T_2^2\right)+\frac{47}{108} \left(T_1^3+T_2^3\right)  \\
&-\frac{413}{192} \left(T_1^2 T_2+T_1 T_2^2\right) -\frac{235 }{288}\left(T_1^4+T_2^4\right)+\frac{9307 }{1536}\left(T_1^3 T_2+T_2^3 T_1\right)\\
&+\frac{2095 }{768} T_1^2 T_2^2+\mathcal{O}(T^5).\ea\ee 
Likewise
\be  F_2^D(T_1,T_2)= -{1\over 240}(T_1^{-2}+T_2^{-2})+\frac{7}{384}  (T_1+T_2)+\frac{31}{24}T_1 T_2-\frac{893}{3456} \left(T_1^2+T_2^2\right)+\mathcal{O}(T^3),\ee 
and similarly for higher genus free energies.

For the $SU(4)$ theory the matrix model \eqref{su3mm} reads
\be \label{caso4} {\ba Z_{\rm 4}^{\rm 4d}(M_1,M_2,M_{3})=&  {1 \over M_1!  M_2! M_{3}!} \int {{\rm d} ^M x\over (2\pi)^M} \prod_{i_1=1}^{M_1} \re^{-  \frac{4 \Lambda }{ \pi ^2 \epsilon}{ \sin \left(\frac{\pi   }{4}\right)} \cosh(x_{i_1})} 
\prod_{i_2=1+M_1}^{M_1+M_2}\re^{-  \frac{4 \Lambda }{ \pi ^2 \epsilon}{ \sin \left(\frac{2 \pi   }{4}\right)} \cosh(x_{i_2})}\\
& \prod_{i_3=1+M_1+M_2}^{M_1+M_2+M_3} \re^{-  \frac{4 \Lambda }{ \pi ^2 \epsilon}{ \sin \left(3 \frac{\pi  }{4}\right)} \cosh(x_{i_3})} 
\\
&   \times {\prod_{ 1\leq i<j \leq M}2\sinh\left({x_i-x_j\over 2}+{1\over 2} (d_i-d_j) \right)2\sinh\left({ x_i-x_j \over 2}+{1\over 2}(f_i-f_j)\right) \over \prod_{i,j=1}^M2\cosh\left( {x_i-x_j \over 2} +{1\over 2} (d_i-f_j) \right)},
\ea }\ee
where
\be  d_{i} =  \left\{\begin{array}{c c c}
-{  \ri \pi  \over 2}    \quad \text{if}\quad  &1 \leq i \leq M_1 ,  \\
-{  \ri \pi  \over 4}  \quad \text{if}\quad  & 1+M_1 \leq i \leq M_1+M_2 , \\
\quad 0 \quad \text{if} \quad  & i > M_1+M_2.  
 \end{array}\ \right.
\ee
 and $f_j=-\frac{1}{2} \ri \pi - d_j$.  The corresponding genus zero free energy reads
 \be \ba 
F_0^D(T_1,T_2,T_3)=&\frac{1}{2} T_1^2 \left(\log \left(\frac{T_1}{2}\right)-\frac{3}{2}\right)+\frac{1}{2} T_3^2 \left(\log \left(\frac{T_3}{2}\right)-\frac{3}{2}\right)+\frac{1}{2} T_2^2 \left(\log \left(\frac{T_2}{4 \sqrt{2}}\right)-\frac{3}{2}\right)\\
&-T_1-\sqrt{2} T_2-T_3-T_3 T_1 \log (2)+T_2 T_1 \log \left(3-2 \sqrt{2}\right)+T_2 T_3 \log \left(3-2 \sqrt{2}\right)\\
&-\frac{1}{2} T_1^3+\sqrt{2} T_2 T_1^2+\frac{1}{4} T_3 T_1^2+T_2^2 T_1+\frac{1}{4} T_3^2 T_1+\sqrt{2} T_2 T_3^2+T_2^2 T_3-\frac{T_3^3}{2}-\frac{T_2^3}{4 \sqrt{2}}\\
&+\frac{3T_1^4}{4}-3 \sqrt{2}T_1^3T_2-\frac{7T_1^3T_3}{16}-\frac{5T_1^2T_2^2}{2}+\frac{T_1^2T_2T_3}{\sqrt{2}}-\frac{5T_1T_2^3}{2 \sqrt{2}}+2T_1T_2^2T_3+\frac{5T_2^4}{64}\\
&-\frac{5T_1^2T_3^2}{16}+\frac{T_1T_2T_3^2}{\sqrt{2}}-\frac{7T_1T_3^3}{16}-\frac{5T_2^3T_3}{2 \sqrt{2}}-\frac{5T_2^2T_3^2}{2}-3 \sqrt{2}T_2T_3^3+\frac{3T_3^4}{4}+\mathcal{O}(T^5)
\ea\ee
 This agrees with \cite{dp1}. 
 From  \eqref{caso4} one can also compute higher genus contributions, for instance \be \ba F_1^D(T_1,T_2,T_3)=&-{1\over 12} \log \left(T_2T_1 T_3\right) + \frac{3 T_1}{8}+\frac{T_2}{8 \sqrt{2}}+\frac{3 T_3}{8}-\frac{11 T_1^2}{16}+\frac{11 T_1 T_2}{4 \sqrt{2}}+\frac{5 T_1 T_3}{16}-\frac{3 T_2^2}{64}\\
 &+\frac{11 T_2 T_3}{4 \sqrt{2}}-T_3^2\frac{11}{16} +\mathcal{O}(T^3).\\
 \ea\ee
 Likwise
  \be \ba F_2^D(T_1,T_2,T_3)=&-{1\over 240}(T_1^{-2}+T_2^{-2}+T_3^{-2})+\frac{5T_1}{128}+\frac{T_2}{256 \sqrt{2}}+\frac{5 T_3}{128}+\mathcal{O}(T^2).\\&\ea\ee
\bibliographystyle{JHEP}

\linespread{0.6}
\bibliography{biblio}
\end{document}